\documentclass[aps]{revtex4}

\usepackage{amsmath,amsfonts}
\usepackage[dvips]{graphicx}

\begin{document}

\title{Top ten accelerating cosmological models}

\author{Marek Szyd{\l}owski}
 \affiliation{Astronomical Observatory, Jagiellonian University, 
Orla 171, 30-244 Krak{\'o}w, Poland}
 \affiliation{Complex Systems Research Centre, Jagiellonian University, 
Reymonta 4, 30-059 Krak{\'o}w, Poland}
\author{Aleksandra Kurek}
 \affiliation{Astronomical Observatory, Jagiellonian University, 
Orla 171, 30-244 Krak{\'o}w, Poland}
\author{Adam Krawiec}
 \affiliation{Institute of Public Affairs, Jagiellonian University, 
Rynek G{\l}{\'o}wny 8, 31-042 Krak{\'o}w, Poland}
 \affiliation{Complex Systems Research Centre, Jagiellonian University, 
Reymonta 4, 30-059 Krak{\'o}w, Poland}

\begin{abstract}
Recent astronomical observations indicate that the Universe is presently almost 
flat and undergoing a period of accelerated expansion. Basing on Einstein's 
general relativity all these observations can be explained by the hypothesis 
of a dark energy component in addition to cold dark matter (CDM). Because the 
nature of this dark energy is unknown, it was proposed some alternative 
scenario to explain the current accelerating Universe. The key point of this 
scenario is to modify the standard FRW equation instead of mysterious dark 
energy component. The standard approach to constrain model parameters, based on 
the likelihood method, gives a best-fit model and confidence ranges for those 
parameters. We always arbitrary choose the set of parameters which define a 
model which we compare with observational data. Because in the generic case, 
the introducing of new parameters improves a fit to the data set, there 
appears the problem of elimination of model parameters which can play an 
insufficient role. The Bayesian information criteria of model selection (the 
AIC and BIC) are dedicated to promotion a set of parameters which should be 
incorporated to the model. We divide class of all accelerating cosmological 
models into two groups according to the two types of explanation acceleration 
of the Universe. Then the Bayesian framework of model selection is used to 
determine the set of parameters which gives preferred fit to the SNIa data. 
We find a few of flat cosmological models which can be recommend by both the 
Bayes factor and Akaike information criterion. 
\end{abstract}

\maketitle

\section{Introduction}

Recent measurements of distant supernovae (SNIa) 
\cite{Riess:1998cb,Perlmutter:1998np} as well as current measurements of cosmic 
microwave background anisotropies \cite{Spergel:2003cb} favor a spatially flat 
Universe filled by cold dark mater (CDM) 
\cite{deBernardis:2000gy,Padmanabhan:2006ag} and a dark energy component of 
unknown origin \cite{Riess:1998cb,Perlmutter:1998np}, which can be modelled 
as a perfect fluid with energy density $\rho_{X}$ and pressure $p_{X}$ 
violating the strong energy condition $\rho_{X} + 3p_{X} > 0$. The combination 
of CMB and SNIa data with other orthogonal measurements, as the HST 
determination of the Hubble parameter, constrain the Universe to be almost 
flat even if we consider the time variation of dark energy equation of state 
\cite{Ichikawa:2005nb}. All these models explaining current acceleration of 
the Universe in terms of smoothly distributed and slowly varying `dark energy' 
are formulated in the context of the standard cosmological picture based on 
general relativity theory. Possible types of explanation include a cosmological 
constant $\Lambda$ \cite{Weinberg:1989}, an evolving scalar field (quintessence) 
\cite{Ratra:1987rm}, the phantom energy \cite{Caldwell:1999ew,Dabrowski:2003jm} 
in which a weak energy condition is violated, models filled with Chaplygin gas 
\cite{Kamenshchik:2001cp}, models with dynamical coefficient equation of state 
$w_{X} \equiv p_{X}/\rho_{X}$ (decaying vacuum energy density), usually 
parameterized by the scale factor $a$ or redshift $z$, where $(1+z=a^{-1})$, 
fluid which describe quantum effects coming from massless scalar field (the 
Casimir effect) \cite{Ishak:2005xp,Szydlowski:1987vr,Singh:2005xg}, the 
noninteracting Chaplygin gas and baryonic matter  
\cite{Biesiada:2004td,Brevik:2005bj,Szydlowski:2006ma}. In Table~\ref{tab:1} 
we collected the ten candidates for dark energy description together with their 
Friedmann first integral 
\begin{equation}\label{eq:1}
3H^{2}= \rho_{\text{eff}} - \frac {3k}{a^{2}},   
\end{equation}
where $H=(d\ln a)/(dt)$ is the Hubble function, and $k=0,\pm 1$ is the 
curvature index). For all these approaches some hypothetical dark energy 
component is postulated which satisfies the conservation condition 
\begin{equation}\label{eq:2}
\dot{\rho}_{i} = -3H(\rho_{i} + p_{i}), \quad i=\text{m},X
\end{equation}
for both standard dust matter as well as dark energy $X$ separately.

\begin{turnpage}
\begin{table}
\caption{The Hubble function for 10 prototypes of cosmological models 
explaining the present acceleration of the Universe in terms of dark energy}
\label{tab:1}
\begin{tabular}{p{0.5cm}p{4cm}p{0.3cm}lp{0.3cm}l}
\hline
case & model && $H^{2}(z)$ relation && independent model parameters\\
\hline 
1&$\Lambda$CDM model& &
 $H^{2}=H^{2}_{0}\{\Omega_{\text{m},0}(1+z)^{3}+(1- \Omega_{\text{m},0})\}$ && 
 $H_{0}$, $\Omega_{\text{m},0}$ \smallskip \\
2&non-flat FRW model with $\Lambda$& &
 $H^{2}=H^{2}_{0}\{\Omega_{\text{m},0}(1+z)^{3}+\Omega_{k,0}(1+z)^{2} + 
 \Omega_{\Lambda,0}\}$ 
 && $H_{0}$, $\Omega_{\text{m},0}$, $\Omega_{\Lambda,0}$ \smallskip \\
3&FRW model with 2D topological defects $p_{X}=-\frac{2}{3} \rho_{X}$& &
 $H^{2}=H^{2}_{0}\{\Omega_{\text{m},0}(1+z)^{3}+\Omega_{k,0}(1+z)^{2} + 
 \Omega_{\text{topo}}(1+z)\}$ 
 && $H_{0}$, $\Omega_{\text{m},0}$, $\Omega_{k,0}$ \smallskip \\
4 &FRW model with phantom dark energy $p_{X}=-\frac{4}{3} \rho_{X}$& &
 $H^{2}=H^{2}_{0}\{\Omega_{\text{m},0}(1+z)^{3}+\Omega_{k,0}(1+z)^{2} + 
 \Omega_{ph,0}(1+z)^{-1}\}$ && 
 $H_{0}$, $\Omega_{\text{m},0}$, $\Omega_{k,0}$ \smallskip \\
5&FRW model with phantom dark energy $p_{X}=w_{X}\rho_{X}$, $w_{X}<-1$ fitted&& 
 $H^{2}=H^{2}_{0}\{\Omega_{\text{m},0}(1+z)^{3}+\Omega_{k,0}(1+z)^{2} + 
 \Omega_{\text{ph},0}(1+z)^{3(1+w_{X})}\}$ &&
 $H_{0}$, $\Omega_{\text{m},0}$, $\Omega_{k,0},\ w_{X}$ \smallskip \\
6 &FRW model with Chaplygin gas $p_{X}=-\frac{A}{\rho_{X}},\ A>0$&&
 $H^{2}=H^{2}_{0} \left\{\Omega_{\text{m},0}(1+z)^{3}+\Omega_{k,0}(1+z)^{2} + 
 \Omega_{\text{Ch},0}[A_{S}+(1-A_{S})(1+z)^{6}]^{ \frac{1}{2}} \right\}$ && 
 $H_{0}$, $\Omega_{\text{m},0}$, $\Omega_{k,0}$, $A_{S}$ \smallskip \\
7& FRW model with generalized Chaplygin gas 
 $p_{X}=-\frac{A}{\rho_{X}^{\alpha}},$ $A>0$, $\alpha = \text{const}$&&
 $H^{2}=H^{2}_{0} \left\{\Omega_{\text{m},0}(1+z)^{3}+
 \Omega_{k,0}(1+z)^{2} + \Omega_{\text{Ch},0}[A_{S}+(1-A_{S})
 (1+z)^{3(1+\alpha)}]^{\frac{1}{1+\alpha}}\right\}$ &&
 $H_{0}$, $\Omega_{\text{m},0}$, $\Omega_{k,0}$, $A_{S}$, $\alpha$ \smallskip \\
8a & FRW models with dynamical E.Q.S. parameterized by $z$ 
 $p_{X}=(w_{0}+w_{1}z)\rho_{X}$&&
 $H^{2}=H^{2}_{0} \left\{\Omega_{\text{m},0}(1+z)^{3}+ \Omega_{k,0}(1+z)^{2} + 
 \Omega_{X,0}(1+z)^{3(w_{0}-w_{1}+1)} \exp [3w_{1}z] \right\}$ && 
 $H_{0}$, $\Omega_{\text{m},0}$, $\Omega_{X,0}$, $w_{0}$, $w_{1}$ \smallskip \\
8b & FRW models with dynamical E.Q.S parameterized by scale factor $a$
 $p_{X}=(w_{0}+$ $w_{1}(1-a))\rho_{X}$&&
 $H^{2}=H^{2}_{0} \left\{\Omega_{\text{m},0}(1+z)^{3}+\Omega_{k,0}(1+z)^{2} + 
 \Omega_{X,0}(1+z)^{3(w_{0}+w_{1}+1)} \exp [- \frac {3w_{1}z}{1+z}] \right\}$ &&
 $H_{0}$, $\Omega_{\text{m},0}$, $\Omega_{X,0}$, $w_{0}$, $w_{1}$ \smallskip \\
9& FRW model with quantum effects origin from massless scalar field at 
 low temperature (Casimir effect) $\rho_{X}=-\frac{\rho_{X,0}}{a^{4}},$ 
 $\rho_{X,0}>0 $&&
 $H^{2}=H^{2}_{0}\{\Omega_{\text{m},0}(1+z)^{3}+\Omega_{k,0}(1+z)^{2} + 
 \Omega_{\Lambda,0} - \Omega_{\text{Cass},0}(1+z)^{4}\}$ &&
 $H_{0}$, $\Omega_{\text{m},0}$, $\Omega_{\Lambda,0}$, $\Omega_{\text{Cass},0}$ 
 \smallskip \\
10&flat FRW model with Chaplygin gas and baryonic matter 
 $p_{\text{eff}}=0-3 \bar{\alpha} \rho^{m}H$; 
 $\Omega_{\text{m},0}=0.05$, $m=-1.5$ are fixed &&
 $H^{2}=H^{2}_{0} \left\{(1-\Omega_{\text{m},0})[A_{S}+
 (1-A_{S}) (1+z)^{3(\frac{1}{2}-m)}]^{\frac{2}{1-2m}} + 
 \Omega_{\text{m}}(1+z)^3 \right\}$ && 
 $H_{0}$, $A_{S}$ \\
\hline
\end{tabular}
\end{table}
\end{turnpage}

On the other hand, alternative ideas to the dark energy idea have been 
recently offered. Freese and Lewis \cite{Freese:2002sq} proposed so called 
the Cardassian model, in which the Universe is flat, matter dominated and 
accelerating not due to dark energy but as a consequence of modification of 
the Friedmann first integral as follows
\begin{equation}\label{eq:3}
3H^{2}=\rho + B \rho ^{n},
\end{equation}
where $B$ is a constant and the 
energy density contain only dust matter and radiation ($8 \pi G =c=1$). The 
additional second term in relation~(\ref{eq:3}) may arise from `new physics'. 
It dominates at the late epoch and drives the present acceleration of the 
Universe. Because terms of type $\rho^{n}$ may arise as a consequence of 
living on $(3+1)$ brane in a high-dimensional bulk space, the origin of 
acceleration lies rather in modification of the FRW equation at low energy 
scales than due to a dark energy component. In brane world scenarios, 
the observer is embedded on 
the brane in a larger space on which gravity can propagate. The idea is that 
an observer measures only $4$-dimensional gravity up to some corrections that 
given the weakness of gravity, can in general be made small enough not to 
conflict with observations without tweaking with a model parameter too much 
\cite{Avelino:2001qh}.

In Table~\ref{tab:2} it is collected ten representative models offering 
explanation of current acceleration of the Universe in an alternative way to 
a dark energy. Apart from the Cardassian model there are different brane world 
scenarios which were originally proposed by Dvali, Gabadadze and Porrati (DGP) 
\cite{Dvali:2000hr}, Deffayet, Dvali and Gabadadze (DDG) \cite{Deffayet:2000uy}, 
Randall and Sundrum \cite{Randall:1999ee}, Shtanov brane models 
\cite{Shtanov:2000vr}. We also consider models of `nonlinear gravity' based 
on modified Lagrangian density which is proportional to $R^{n}$, where $R$ is 
the Ricci scalar and $n$ is constant \cite{Borowiec:2006hk} ($n=1$ for 
standard general relativity), models based on non-Riemannian gravity 
\cite{Krawiec:2005jj}, bouncing models arising in the context of loop quantum 
gravity and models with energy transfer between the dark matter and dark 
energy sectors ($\Lambda$ decaying vacuum and phantom dark energy).

\begin{turnpage}
\begin{table}
\caption{The Hubble function for 10 cosmological models beyond 
the standard general relativity} 
\label{tab:2}
\begin{tabular}{p{0.5cm}p{4cm}p{0.3cm}lp{0.3cm}l}
\hline
 case & model && $H^{2}(z)$ relation && independent model parameters\\
\hline 
1&Cardassian type of Friedmann equation, $\Omega_{r,0}=10^{-4}$ is fixed && 
 $H^{2}=H^{2}_{0} \left\{ \Omega_{k,0}(1+z)^{2}+\Omega_{\text{m},0}(1+z)^{4} 
 \left[ \frac{1}{1+z} + (1+z)^{-4+4n} \left ( \frac{1- \Omega_{r,0}-
 \Omega_{\text{m},0}}{\Omega_{\text{m},0}} \right ) \left ( \frac{ \frac{1}{1+z}+
 \frac{\Omega_{r,0}}{\Omega_{\text{m},0} }}{ 1+ 
 \frac{\Omega_{r,0}}{\Omega{m,0}}}\right)^{n} \right ] \right\}$ &&
 $H_{0},\ \Omega_{k,0},\ n$ \smallskip \\
2&Dvali-Gabadadze-Porrati brane models (DGP) &&
 $H^{2}=H^{2}_{0} \{ \left [ \sqrt{\Omega_{\text{m},0}(1+z)^{3}+\Omega_{rc,0}}+
 \sqrt{\Omega_{rc,0}} \right] ^{2} + \Omega_{k,0}(1+z)^{2} \}$ &&
 $H_{0}$, $\Omega_{\text{m},0}$, $\Omega_{rc,0}$ \smallskip \\
3&Deffayet-Dvali-Gabadadze brane models with $\lambda$ (DDG) &&
 $H^{2}=H^{2}_{0} \left\{ \left [ - \frac{1}{2r_{0}H_{0}} +
 \sqrt{\Omega_{\text{m},0}(1+z)^{3}+\Omega_{\lambda,0} + 
 \frac{1}{4r_{0}^{2}H_{0}^{2}}}\right]^{2}\right\}$ &&
 $H_{0}$, $r_{0}$, $\Omega_{\lambda,0}$ \smallskip \\
4&Randall-Sundrum brane models with dark radiation and $\Lambda=0$ &&
 $H^{2}=H^{2}_{0} \{ \Omega_{\text{m},0}(1+z)^{3}+\Omega_{k,0}(1+z)^{2}+ 
 \Omega_{dr,0}(1+z)^{4}+ \Omega_{\lambda,0}(1+z)^{6} \}$ &&
 $H_{0}$, $\Omega_{\text{m},0}$, $\Omega_{dr,0}$, $\Omega_{\lambda,0}$ 
 \smallskip \\
5&Randall-Sundrum brane models with dark radiation and $\Lambda$ (RSB) &&
 $H^{2}=H^{2}_{0} \{ \Omega_{\text{m},0}(1+z)^{3}+\Omega_{k,0}(1+z)^{2}+ 
 \Omega_{dr,0}(1+z)^{4}+ \Omega_{\lambda,0}(1+z)^{6} + \Omega_{\Lambda,0}\}$ &&
 $H_{0}$, $\Omega_{\text{m},0}$, $\Omega_{\text{dr},0}$, $\Omega_{\lambda,0}$, 
 $\Omega_{\Lambda,0}$ \smallskip \\
6a&Shtanov brane models (Brane1) &&
 $H^{2}=H^{2}_{0} \left\{ \Omega_{\text{m},0}(1+z)^{3}+\Omega_{\sigma,0}+ 
 2 \Omega_{l,0}-2\sqrt{\Omega_{l,0}}\sqrt{\Omega_{\text{m},0}(1+z)^{3}+
 \Omega_{\sigma,0}+ \Omega_{l,0}+ \Omega_{\Lambda b,0}} \right\}$ &&
 $H_{0}$, $\Omega_{\text{m},0}$, $\Omega_{\sigma,0}$, $\Omega_{l,0}$ 
 \smallskip \\
6b.&Shtanov brane models (Brane2)&& 
 $H^{2}=H^{2}_{0} \left\{ \Omega_{\text{m},0}(1+z)^{3}+\Omega_{\sigma,0}+ 
 2 \Omega_{l,0}+2\sqrt{\Omega_{l,0}}\sqrt{\Omega_{\text{m},0}(1+z)^{3}+
 \Omega_{\sigma,0}+ \Omega_{l,0}+ \Omega_{\Lambda b,0}} \right\}$&&
 $H_{0},\ \Omega_{\text{m},0},\ \Omega_{\sigma,0}$, $ \Omega_{l,0}$ 
 \smallskip \\
7&modified affine gravity (MAG) model&&
 $H^{2}=H^{2}_{0} \{ \Omega_{\text{m},0}(1+z)^{3}+\Omega_{k,0}(1+z)^{2}+ 
 \Omega_{\psi,0}(1+z)^{6}+ \Omega_{\Lambda,0}\}$&&
 $H_{0}$, $\Omega_{\text{m},0}$, $\Omega_{\psi,0}$, $\Omega_{\Lambda,0}$ 
 \smallskip \\
8&FRW models of nonlinear gravity with Lagrangian density proportional to
 Ricci scalar $R$ (NG) &&
 $H^{2}=H^{2}_{0} \{ \Omega_{\text{m},0}(1+z)^{3}\frac{2n}{3-n}+
 \Omega_{r,0}(1+z)^{4} \frac{4n(2-n)}{(n-3)^{2}}\}
 \Omega_{\text{nonl},0}(1+z)^{\frac{3(1-n)}{n}}$&&
 $H_{0}$, $\Omega_{\text{m},0}$, $\Omega_{\text{nonl},0}$, $n$ \smallskip \\
9&bouncing models with $\Lambda$ (B$\Lambda$CDM) &&
 $H^{2}=H^{2}_{0} \{ \Omega_{\text{m},0}(1+z)^{3}+\Omega_{k,0}(1+z)^{2}- 
 \Omega_{n,0}(1+z)^{n}+ \Omega_{\Lambda,0}\}$&&
 $H_{0}$, $\Omega_{\text{m},0}$, $\Omega_{n,0}$, $\Omega_{\Lambda,0}$
 \smallskip \\
10a&models with energy transfer (dark matter $\leftrightarrow$ vacuum energy 
  $\rho_{1}$ sector)&&
  $H^{2}=H^{2}_{0} \{ \Omega_{\text{m},0}(1+z)^{3}+\Omega_{\text{int}}(1+z)^{n}+ 
  \Omega_{\Lambda,0}\}$&&
  $H_{0},\ \Omega_{\text{m},0},\ \Omega_{\Lambda,0},$ \smallskip \\
10b&models with energy transfer (dark matter $\leftrightarrow$ phantom dark 
  energy sector) &&
  $H^{2}=H^{2}_{0} \left\{ \Omega_{\text{m},0}(1+z)^{3}+\Omega_{\text{int}}(1+z)^{n}+ 
  \Omega_{\text{ph}}(1+z)^{3(1+w_{X})} \right\}$ &&
  $H_{0}$, $\Omega_{\text{m},0}$, $\Omega_{\text{ph},0}$, $w_{X}$ \\
  & && && \\
\hline
\end{tabular}
\end{table}
\end{turnpage}

The main goal of this paper is to make the ranking of accelerating models using 
the Bayesian framework \cite{Jeffreys:1935,Jeffreys:1961}. The effectiveness 
of application of information criteria of model selection in the cosmological 
context was recently demonstrated by many authors \cite{Liddle:2004nh,%
John:2002gg,Saini:2003wq,Parkinson:2004yx,Mukherjee:2005tr,Beltran:2005xd,%
Mukherjee:2005wg,Szydlowski:2005xv,Godlowski:2005tw,Eriksen:2006pn,%
Drell:1999dx,Marshall:2004zd,Hobson:2002de,Loredo:2001rx,Kunz:2006mc,%
Trotta:2005ar,Edmondson:2006fy}. 
In the paper we obtain these accelerating models which are the best ones from 
the set under consideration in explaining the SNIa data. We calculate the Bayes 
factor for all models with different numbers of parameters, which differentiate 
between them \cite{Kass:1995}. The Bayes factor measures the change in relative 
probabilities of any two models in light of observational data (SNIa data) 
when we update the prior relative model probabilities to the posterior relative 
model probabilities. 

In observational cosmology many theoretically allowed models with different 
prediction of the past (big bang versus bounce) and the future (big rip versus 
de Sitter attractor) becomes in good agreements with the observational data 
\cite{Szydlowski:2005qb}. Therefore we propose to use the Bayesian factor to 
differentiate among all these models and make some ranking.

\section{Idea of model selection}

In the development of cosmology the basic role played an idea of cosmological 
models together with an idea of astronomical tests \cite{Ellis:1999sx}. The 
idea of cosmological tests make cosmological models parts of astronomy which 
offers possibility of observationally determining the set of realistic 
parameters, that can characterize viable models. While we can perform 
estimation of model parameters using a standard minimization procedure based 
on the likelihood method, many different scenarios are still in a good 
agreement with observational data of SNIa. This problem which appears in 
observational cosmology is called the degeneracy problem. To solve this 
problem it is required to differentiate between different dark energy models. 
We propose to use the Akaike information criterion (AIC) and BIC quantity (as 
an approximation to the logarithm of the marginal likelihood).  

For the model selection framework it is required to have data and a set of 
models and then we can make the model based statistical inference. The model selection 
should be based on a well-justified single (even naive) model or, at least, a 
simple model which suffices for making inferences from the data. In our case 
the $\Lambda$CDM model plays just the role of such a model. The model selection 
should be viewed as a way to obtain model weights, not just a way to select 
only one model (and then ignore that selection occurred). Moreover in the 
notion of true models do not believe information theories because the model 
by definition is only approximations to unknown physical reality: there is no 
true model of the Universe that perfectly reflect large structure of space-time, 
but some of them are useful.

In this paper the Bayes factor and AIC quantity are used to compare models 
gathered in Table~\ref{tab:1} and Table~\ref{tab:2}. 

The AIC is defined in the following way \cite{Akaike:1974}
\begin{equation}\label{eq:4}
\text{AIC}=-2\ln\mathcal{L} + 2d,
\end{equation}
where $\mathcal{L}$ is the maximum likelihood and $d$ is a number of model 
parameters. The best model with a parameter set providing the preferred fit to 
the data is that which minimizes the AIC. While there are justification of the 
AIC in information theory and also rigorous statistical foundation for the AIC, 
it can be also justified as Bayesian using a `savvy' prior on models that is a 
function of a sample size and a number of model parameters. For the AIC we can 
define $\Delta$AIC$_{i}$ as the difference of the AIC for model $i$ and the AIC 
value for the reference model: $\Delta_{i}=\text{AIC}_{i} - \text{AIC}_{1}$. 
The $\Delta_{i}$ are easy to interpret and allow a quick `strength of evidence' 
comparison and a ranking of candidates for dark energy description. The models 
with $0 \le \Delta_{i}\le 2$ have substantial support (evidence), those where 
$4<\Delta_{i}\le 7$ have considerably less support, while models having 
$\Delta_{i} > 10 $ have essentially no support with respect to model 1.

In the Bayesian framework a best model (from the model set $\{M_{i}\}$, 
$i=1,\dots,K$) is that which has the largest value of probability in the light 
of data (so called a posteriori probability). We can define the posterior odds 
for models $M_{i}$ and $M_{j}$, which (in the case when no model is favored a 
priori) is reduced to the marginal likelihood ($E$) ratio (so called the Bayes 
factor -- $B_{ij}$)
\begin{equation}\label{eq:7}
B_{ij}=\frac{\int L(\bar{\theta}|D,M_{i})P(\bar{\theta}|M_{i}) 
d \bar{\theta} }{\int L(\bar{\eta}|D,M_{j})P(\bar{\eta}|M_{j}) 
d \bar{\eta}}=\frac{E_{i}}{E_{j}},
\end{equation}
where $\bar{\theta}$ is a parameter vector, which defines model $i$, 
$L(\bar{\theta}|D,M_{i})$ is likelihood under model $i$, 
$P(\bar{\theta}|M_{i})$ is the prior probability for ${\bar{\theta}}$ under 
a model $i$.

Schwarz \cite{Schwarz:1978} showed that for iid observations coming from the 
linear exponential family distribution the asymptotic approximation 
($N\to \infty$) to the logarithm of the marginal likelihood is given by
\begin{equation}\label{eq:9}
\ln E = \ln \mathcal{L} - \frac{d}{2}\ln N + O(1).
\end{equation}

According to this result he introduced a criterion for the model 
selection: the best model is that which minimized the BIC, defined as
\begin{equation}\label{eq:11}
\text{BIC}=-2 \ln \mathcal{L} + d \ln N.
\end{equation}

This criterion can be derived in in such a way that it is not required to 
assume any specific form for the likelihood function but it is only 
necessary that the likelihood function satisfies some non-restrictive 
regularity conditions. Moreover the data do not need to be independent 
and identically distributed \cite{Cavanaugh:1999}.

To compare models $M_{i}$ and $M_{j}$ one can compute $2 \ln B_{ij} = 
- (\text{BIC}_{i} - \text{BIC}_{j}) \equiv - \Delta \text{BIC} _{ij}$ which can 
be interpret as `strength of evidence' against $j$ model: $0\leq 2\ln B_{ij} < 
2$--not worth more than a bare mention, $2\leq 2\ln B_{ij} < 6$ -- positive, 
$6\leq 2\ln B_{ij} < 10$ -- strong, and $2\ln B_{ij}\ge 10$ -- very strong.

It is useful to choose one model from our model set (a reference model--$s$) and 
compare the rest models with this one model, situation in which $2\ln B_{si}>0$ 
indicates evidence against model $i$ with respect to the reference model, 
whereas $2\ln B_{si}<0$ denotes evidence in favor of model $i$.

We can compute posterior probability for model $i$ in the following way 
\begin{equation}\label{eq:12}
P(M_{i}|D)=\frac{B_{is}}{\sum _{k=1}^{K} B_{ks}}, 
\end{equation}
where $B_{is}=\exp[\frac{1}{2}\Delta BIC_{si}]$. Then one can see how prior 
believe about model probability $P(M_{i})=\frac{1}{K}$ change after inclusion 
data to analysis. This is the probability for model $i$ being the best model 
from set of models under consideration.  

There are many simulation studies in the statistical literature on either the 
AIC or BIC alone, or often comparing them and making recommendation on which 
one to use. It should be pointed out that both of them are an asymptotic 
approximation. This assumption is satisfied when sample size used in analysis 
is large enough, large with respect to the number of unknown model parameters.

Note that the assumptions of using model selection methods are satisfied, namely
\begin{enumerate}
\item there is model-based inference from SNIa data (the luminosity distance 
observable);
\item there is a set of models and no certainty about which model should be 
used in explanation of present acceleration;
\item a data-based choice is made among these competing models (see 
Table~\ref{tab:1}, Table~\ref{tab:2}).
\end{enumerate}

In this paper the restricted `Gold' sample of $N=157$ SNIa \cite{Riess:1998cb} 
is used. It is assumed that the supernovae measurements come with uncorrelated 
Gaussian errors. In this case the likelihood function has the following form
\begin{equation}
L \propto \exp \left[-\frac{1}{2}\left(\sum_{i=1}^{N}
\frac{(\mu_{i}^{\text{theor}}-\mu_{i}^{\text{obs}})^{2}}{\sigma_{i}^{2}}\right) 
\right],
\end{equation}
where $\sigma_{i}$ is known, $\mu_{i}^{\text{obs}}=m_{i}-M$ ($m_{i}$--the 
apparent magnitude, $M$--the absolute magnitude of SNIa), 
$\mu_{i}^{\text{theor}}=5\log_{10}D_{Li} + \mathcal{M}$, 
$\mathcal{M}=-5\log_{10}H_{0}+25$ and $D_{Li}=H_{0}d_{Li}$, 
where $d_{Li}$ is luminosity distance
\begin{equation}\label{eq:16}
d_{Li}=(1+z_{i}) \frac{c}{H_{0}} \frac{1}{\sqrt{|\Omega_{k,0}|}}
\mathcal{F} \left( H_0 \sqrt{|\Omega_{k,0}|} \int_{0}^{z_{i}} 
\frac{d z'}{H(z')} \right),
\end{equation}
where $\Omega_{k,0} = - \frac{k}{H_0^2a_{0}^{2}}$ and 
\begin{align*}
\mathcal{F}(x) &= \sinh(x) &\text{for} & &k<0\\ 
\mathcal{F}(x) &= x        &\text{for} & &k=0\\
\mathcal{F}(x) &= \sin(x)  &\text{for} & &k>0.\\
\end{align*}

Table~\ref{tab:3} gives the value of the AIC and BIC for flat models from 
Table~\ref{tab:1}. It also contains values of $\Delta\text{AIC}_{is}$ and 
$2\ln B_{si}$, where model $s$ is our reference model: the $\Lambda$CDM model 
(indexed by $1$) in the first case and model which minimize both the AIC and 
BIC quantities -- the flat FRW model with noninteracting the Chaplygin gas and 
baryonic matter (indexed 
by $10$) in the second one. These values are also illustrated in Figures~1A, 
1B and Figures~2A, 2B respectively. Table~\ref{tab:4} contains the same 
quantities for flat models from Table~\ref{tab:2}. Here the first reference model 
is that which minimizes the AIC quantity -- the Cardassian model (indexed by $1$), 
the second one is that which minimizes the BIC quantity -- the DGB model (indexed 
by $2$). Note that in this case these models are different (on the contrary 
with situation in Table~\ref{tab:1}, where the same model minimizes both the 
AIC and BIC indices). Figures~3A, 3B and Figures~4A, 4B illustrate values of 
$\Delta\text{AIC}_{is}$ and $2\ln B_{si}$ respectively. Additionally we compare 
all flat models from Table~\ref{tab:2} with the $\Lambda$CDM model (those 
comparisons are illustrated on Figures 5A and 5B.) 
 
\begin{table}
\caption{Values of the AIC, BIC, $\Delta\text{AIC}_{is}$ and $2\ln B_{si}$ 
(where $s$ is the index of the reference model) for the flat models from 
Table~\ref{tab:1}.}
\label{tab:3}
\begin{tabular}{cccr@{.}lr@{.}lr@{.}lr@{.}l}
\hline
  case & AIC & BIC & \multicolumn{2}{c}{$\Delta\textrm{AIC}_{i1}$} & 
  \multicolumn{2}{c}{$2 \ln B _{1i}$} & 
  \multicolumn{2}{c}{$\Delta\textrm{AIC}_{i10}$} & 
  \multicolumn{2}{c}{$2 \ln B_{10i}$} \\  
\hline
 1& 179.9 & $\mathbf{186.0}$ & 0&0&0&0 &2&0 & 2&1  \\
 2& 179.9 & 189.0 & 0&0 & 3&0 & 2&0 &5&1 \\
 3& 183.2 & 189.4 & 3&3 & 3&4 & 5&3 & 5&5 \\
 4& $\mathbf{178.0}$ & $\mathbf{184.1}$ & $-1$&9 & $-1$&9 & 0&1 & 0&2 \\
 5& $\mathbf{178.5}$ & 187.7 & $-1$&4 & 1&7 & 0&6 & 3&8 \\
 6& 179.7 & 188.9 & $-0$&2 & 2&9 & 1&8 & 5&0 \\
 7& 181.7 & 193.9 & 1&8 & 7&9 & 3&8 & 10&0 \\
 8a& 180.5 & 192.7 & 0&6 & 6&7 &2&6 & 8&8 \\
 8b& 180.4 & 192.6 & 0&5 & 6&6 &2&5 & 8&7 \\
 9& 181.9 & 197.1 & 2&0 & 11&1 &4&0 & 13&2 \\
 10& $\mathbf{177.9}$ &$\mathbf{183.9}$ & $-2$&0 & $-2$&1 &0&0 &0&0\\
 \hline
\end{tabular}
\end{table}

\begin{table}
\caption{Values of the AIC, BIC, $\Delta\text{AIC}_{is}$ and $2\ln B_{si}$ 
(where $s$ is the index of the reference model) for the flat models from 
Table~\ref{tab:2}.}
\label{tab:4}
\begin{tabular}{cccr@{.}lr@{.}lr@{.}lr@{.}l}
\hline
  case & AIC & BIC & \multicolumn{2}{c}{$\Delta\text{AIC}_{i1}$} &
  \multicolumn{2}{c}{$2 \ln B_{1i}$} &
  \multicolumn{2}{c}{$\Delta\text{AIC}_{i2}$} &
  \multicolumn{2}{c}{$2 \ln B_{2i}$} \\  
\hline
 1& $\mathbf{178.5}$ & $\mathbf{187.7}$ & 0&0&0&0 & $-2$&4 & 0&7 \\
 2& 180.9 &$\mathbf{187.0}$ & 2&4 & $-0$&7 & 0&0 & 0&0\\
 3& 180.8 & 189.9 & 2&3 & 2&2 & $-0$&1 & 2&9 \\
 4& 327.0 & 336.2 & 148&5 & 148&5 & 146&1 & 149&2 \\
 5& 182.1 & 194.3 & 3&6 & 6&6 & 1&2 & 7&3 \\
 6a& 180.3 & 192.5 & 1&8 & 4&8 & $-0$&6 & 5&5 \\
 6b& 183.8 & 196.0 & 5&3 & 8&3 & 2&9 & 9&0\\
 7& 180.3 & $\mathbf{189.4}$ & 1&8 & 1&7 & $-0$&6 & 2&4\\
 8& 186.6 & 195.8 & 8&1 & 8&1 & 5&7 & 8&8 \\
 9& 183.9 & 196.2 & 5&4 & 8&5 & 3&0 & 9&2 \\
10a& $\mathbf{180.1}$ &192.3 & 1&6 & 4&6 & $-0$&8 & 5&3 \\
10b& $\mathbf{180.2}$ &192.4 & 1&7 & 4&7 & $-0$&7 & 5&4 \\
 \hline
\end{tabular}
\end{table}
 
\begin{figure}
  \includegraphics[height=.26\textheight]{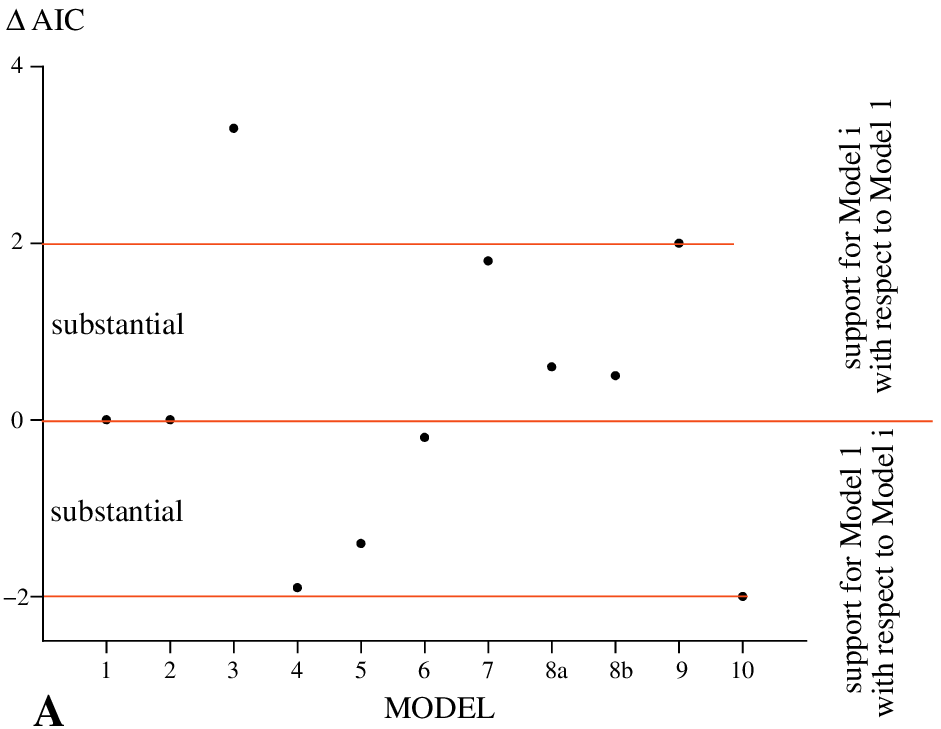}
  \includegraphics[height=.26\textheight]{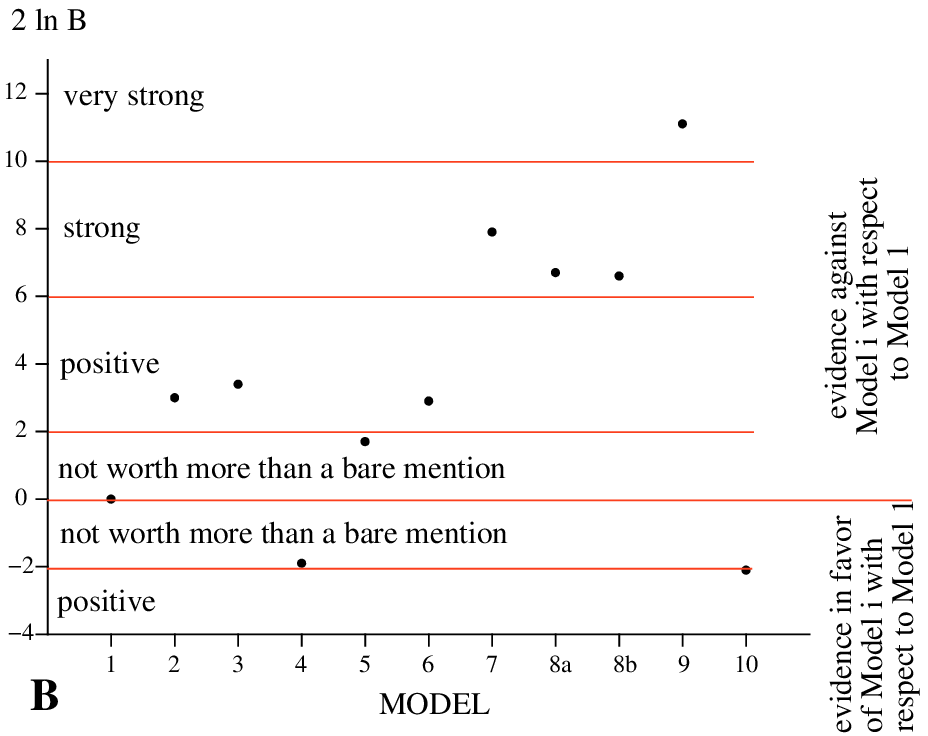}
\caption{Value of A) $\Delta\text{AIC}_{i1}= \text{AIC}_{i}-
\text{AIC}_{1}$ and B) $2\ln B_{1i}$  for models from Table~\ref{tab:1}.}
\end{figure}

\begin{figure}
  \includegraphics[height=.26\textheight]{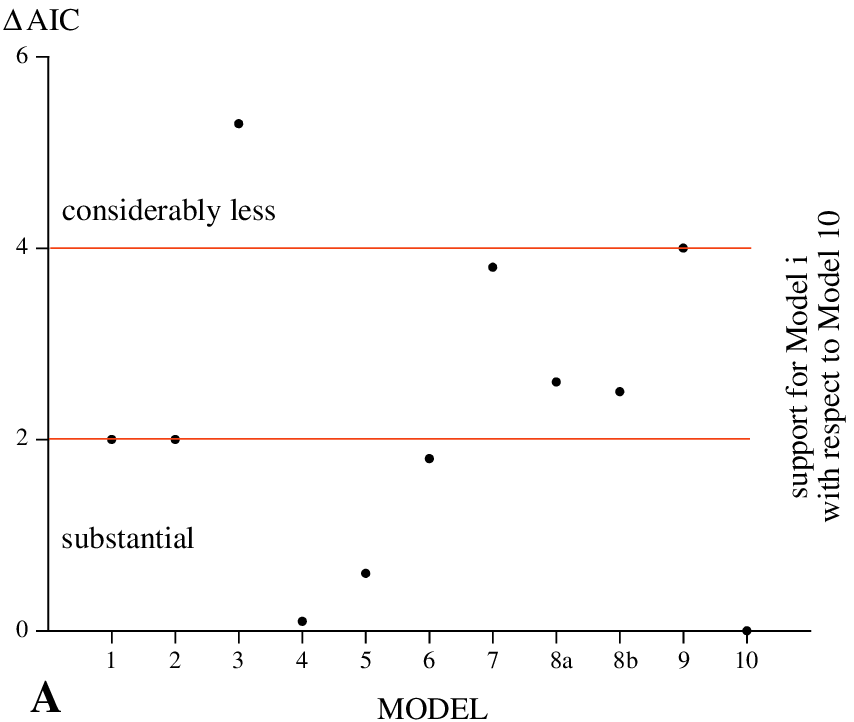}
  \includegraphics[height=.26\textheight]{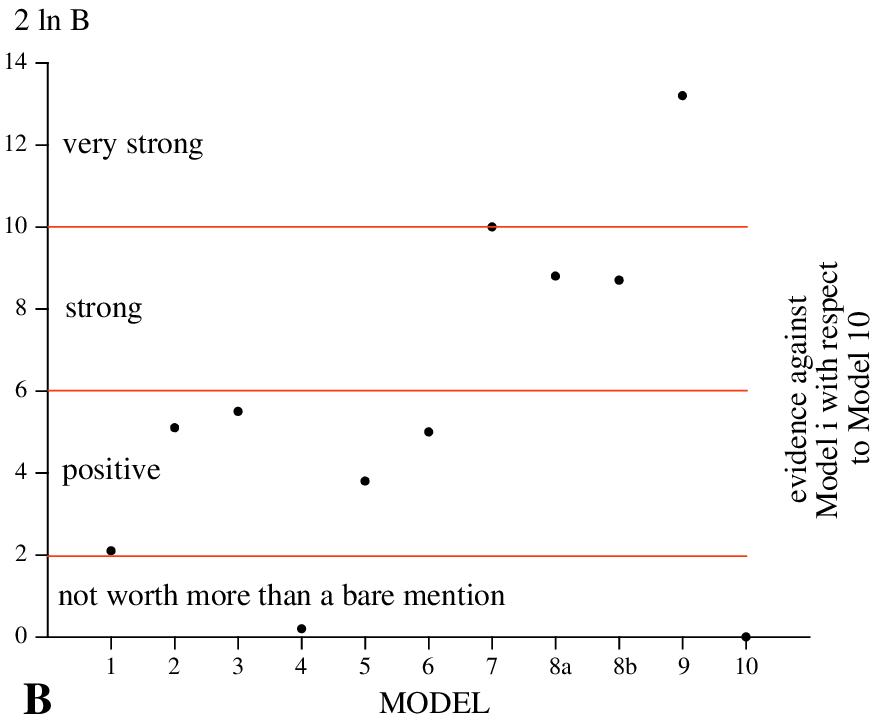}
\caption{Value of A) $\Delta\text{AIC}_{i10}= \text{AIC}_{i}-
\text{AIC}_{10}$ and B) $2\ln B_{10i}$  for models from Table~\ref{tab:1}.}
\end{figure}

\begin{figure}
  \includegraphics[height=.26\textheight]{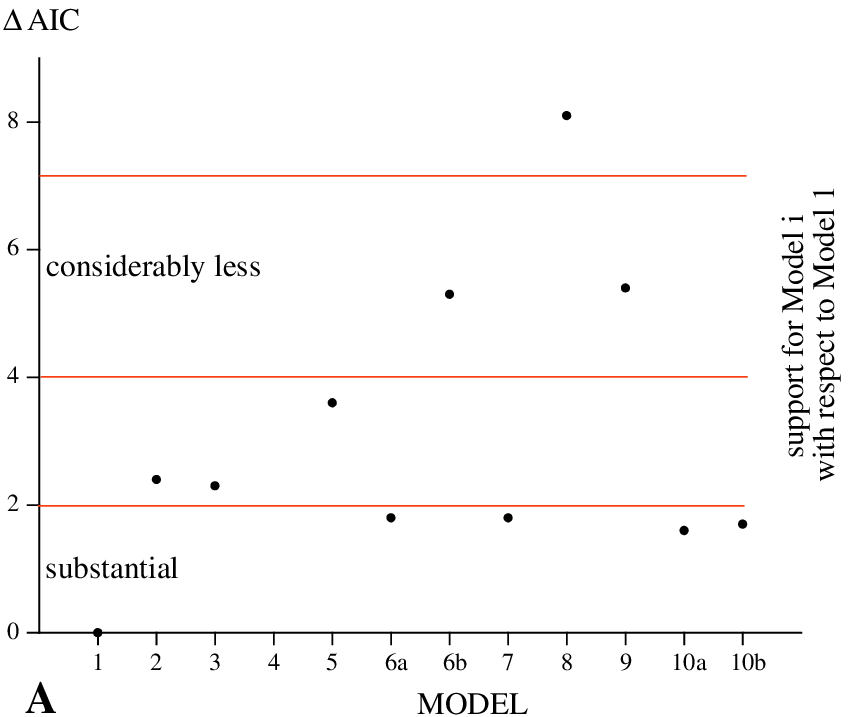}
  \includegraphics[height=.26\textheight]{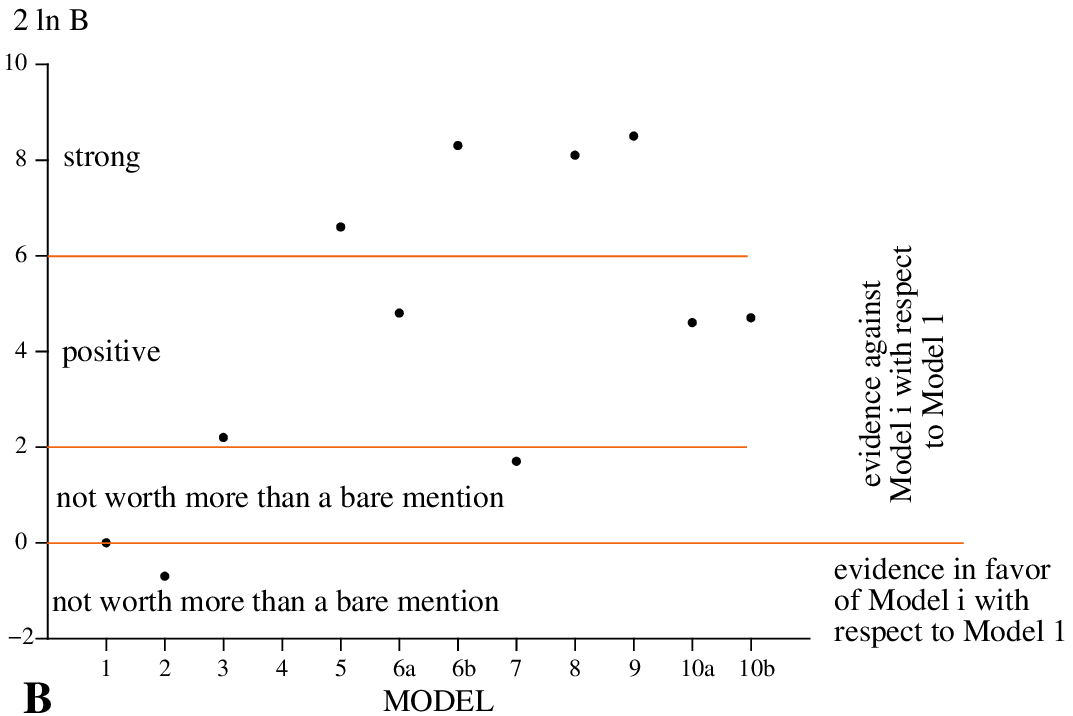}
\caption{Value of A) $\Delta\text{AIC}_{i1}= \text{AIC}_{i}-
\text{AIC}_{1}$ and B) $2\ln B_{1i}$ for models from Table~\ref{tab:2}.}
\end{figure}

\begin{figure}
  \includegraphics[height=.26\textheight]{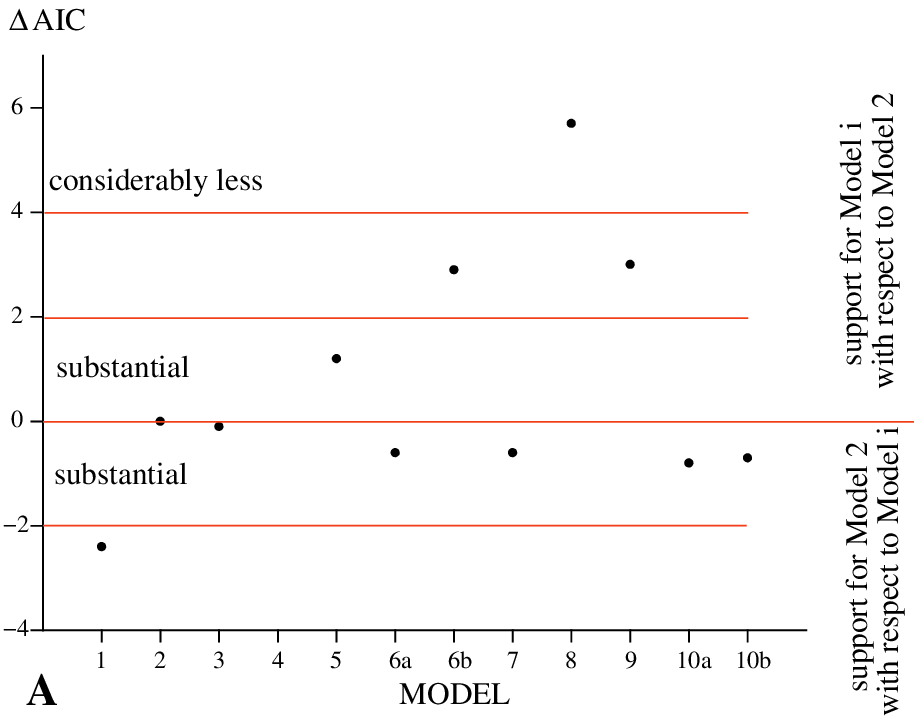}
  \includegraphics[height=.26\textheight]{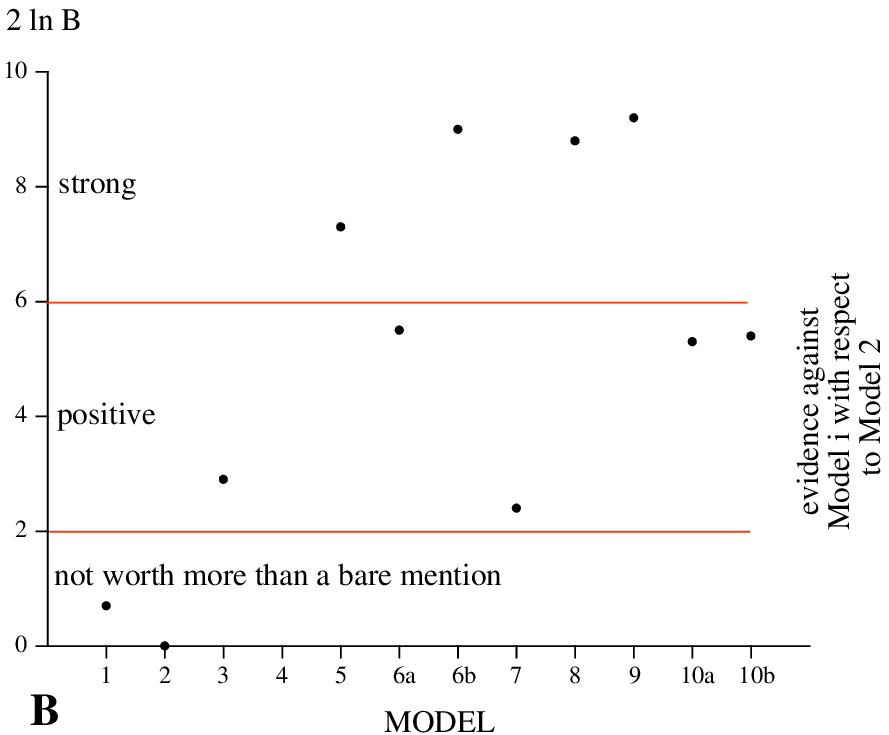}
\caption{Value of A) $\Delta\text{AIC}_{i2}= \text{AIC}_{i}-
\text{AIC}_{2}$ and B) $2\ln B_{2i}$ for models from Table~\ref{tab:2}.}
\end{figure}

\begin{figure}
  \includegraphics[height=.26\textheight]{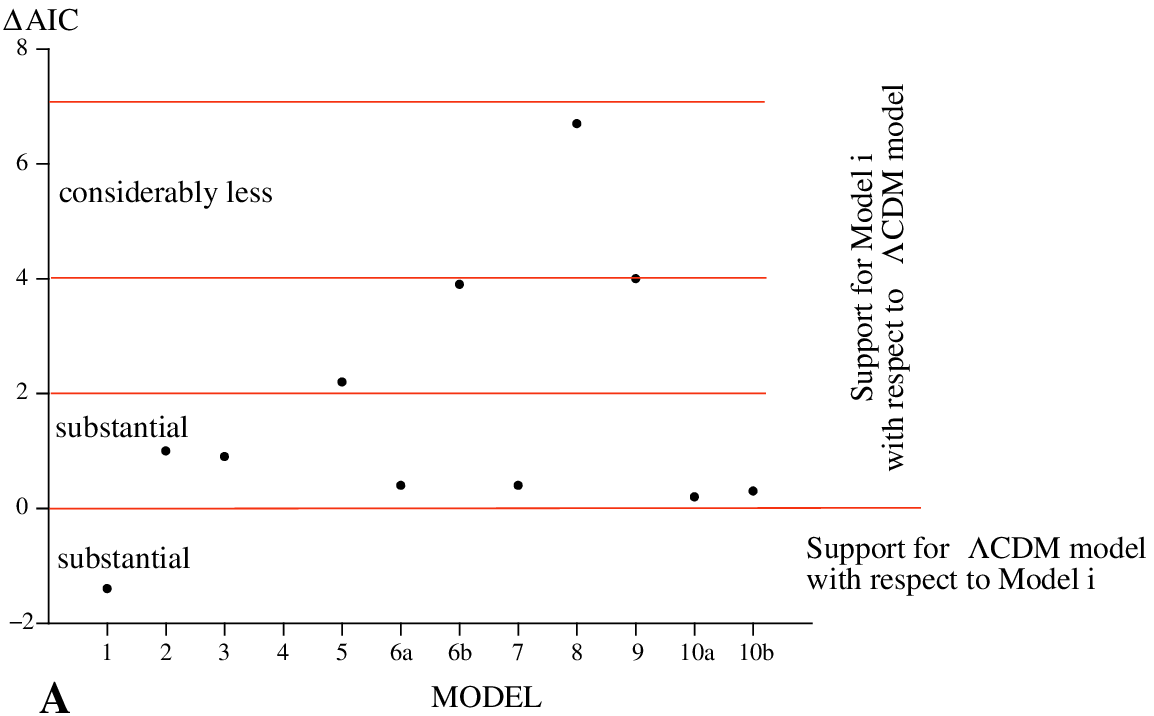}
  \includegraphics[height=.26\textheight]{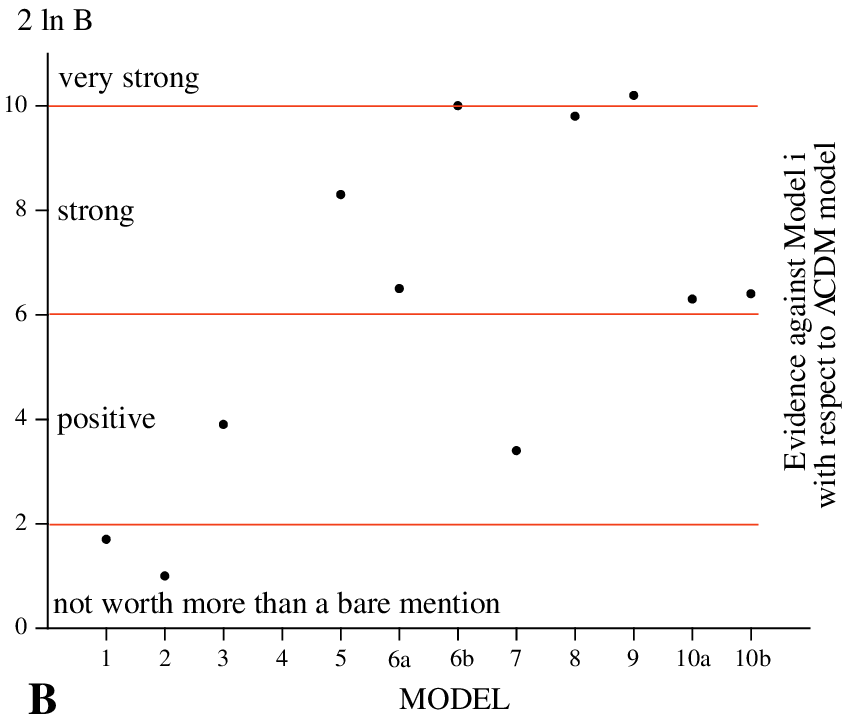}
\caption{Value of A) $\Delta\text{AIC}_{is}= \text{AIC}_{i}-
\text{AIC}_{s}$ and B) $2\ln B_{si}$ for models from Table~\ref{tab:2}, $s$--the 
index of the reference model--the $\Lambda$CDM model.}
\end{figure}

In Figures~6A, 6B are presented values of $\Delta\text{AIC}_{is_{\text{AIC}}}$ 
and $2\ln B_{s_{\text{BIC}}i}$ respectively for set of brane models from 
Table~\ref{tab:2}: $\{2, 3, 4, 5, 6a, 6b\}$. Here the reference model for the AIC 
analysis is the Shtanov Brane1 model (indexed by $6a$), which minimizes the 
AIC quantity in the set under consideration, whereas the DGB model (indexed by 
$2$) for the BIC analysis (this one minimizes the BIC quantity in the models set).

\begin{figure}
  \includegraphics[height=.26\textheight]{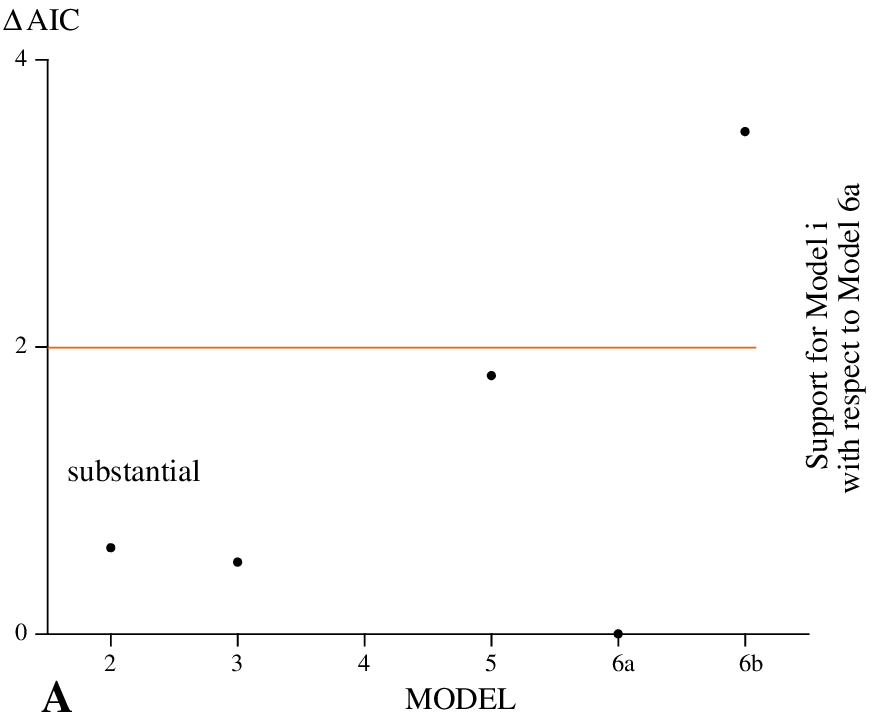}
  \includegraphics[height=.26\textheight]{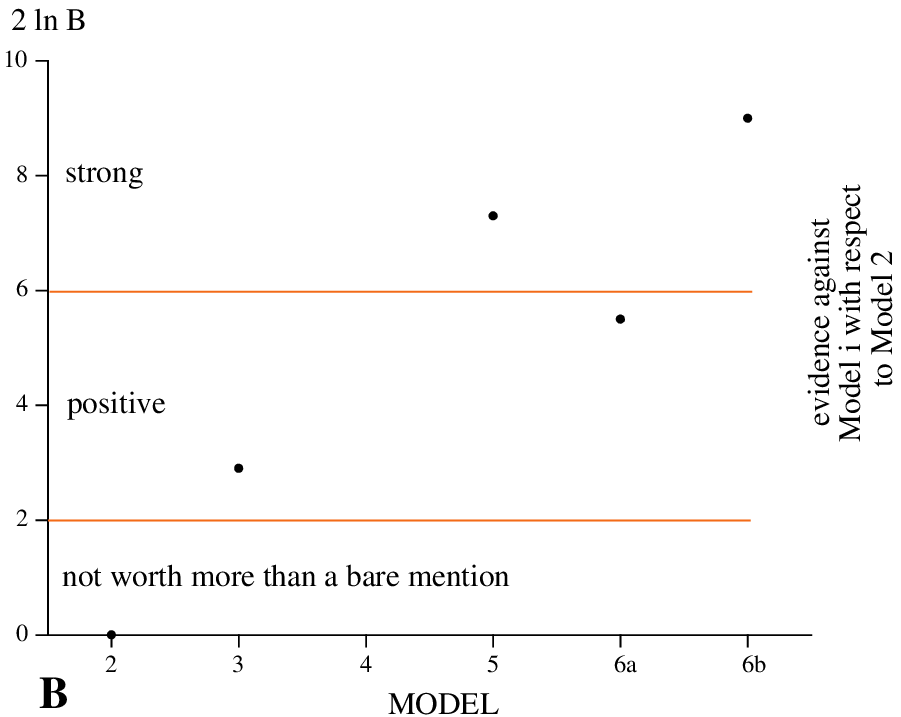}
\caption{Value of A) $\Delta\text{AIC}_{i6a}= \text{AIC}_{i}-
\text{AIC}_{6a}$ and B) $2\ln B_{2i}$  for set of brane models from 
Table~\ref{tab:2}.}
\end{figure}

In Figure~7A are illustrated values of $\Delta\text{AIC}_{is}$ for models 
which have substantial support with respect to the reference model in Figures~2A and 
Figure 3A: set of models from Table~\ref{tab:1}: $\{1, 2, 4, 5, 6, 10\}$ together with 
set of models from Table~\ref{tab:2}: $\{1, 6a, 7, 10a, 10b\}$. Here the reference model 
is this which has a minimal value of the AIC quantity in the set of models under 
consideration -- the flat FRW model with noninteracting Chaplygin gas and baryonic matter (model 
indexed by $10$ from Table~\ref{tab:1}). In Figure~7B are presented values of 
$2\ln B_{si}$ for models from Figures~2B and 4B for which evidence against 
them with respect to the reference model is not worth more than a bare mention: 
Table~\ref{tab:1}: $\{1, 4, 10\}$ and Table~\ref{tab:2}: $\{1, 2\}$. Here the reference 
model is this which minimizes the BIC quantity -- the model indexed by $10$ from 
Table~\ref{tab:1}.

\begin{figure}
  \includegraphics[height=.26\textheight]{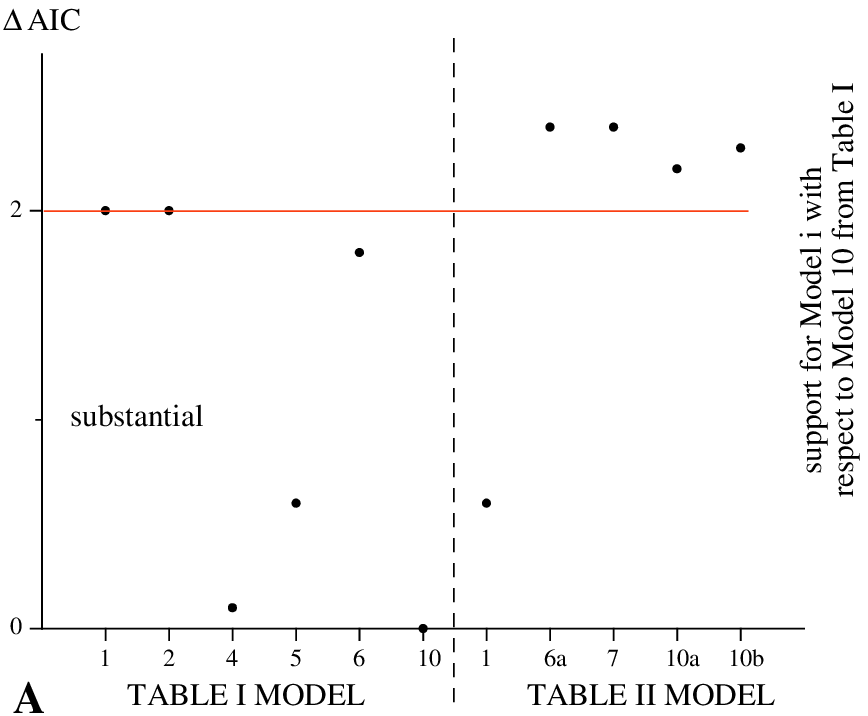}
  \includegraphics[height=.26\textheight]{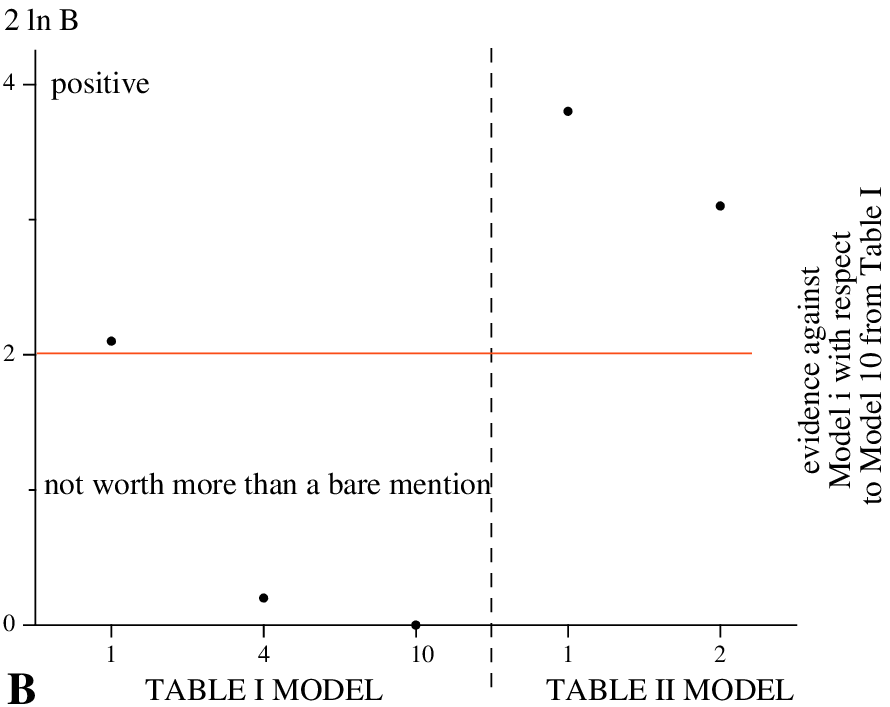}
\caption {Value of A) $\Delta\text{AIC}_{is}= \text{AIC}_{i}-\text{AIC}_{s}$ 
for the set of models from Table~\ref{tab:1}: $\{1, 2, 4, 5, 6, 10\}$ and 
from Table~\ref{tab:2}: $\{1, 6a, 7, 10a, 10b\}$, B) $2\ln B_{si}$ for the 
set of models from Table~\ref{tab:1}: $\{1, 4, 10\}$ and Table~\ref{tab:2}: 
$\{1, 2\}$, $s$--the index of the reference model--the model $10$ in both 
cases.}
\end{figure}
  	
In Tables~\ref{tab:5}, \ref{tab:6}, \ref{tab:7} and \ref{tab:8} are gathered 
values of prior and posterior probabilities (see (\ref{eq:12})) for models 
from Table~\ref{tab:1}, Table~\ref{tab:2}, set of brane models from 
Table~\ref{tab:2}, set of the best models from Table~\ref{tab:1} and 
Table~\ref{tab:2} respectively.

\begin{table}
\caption{Values of the prior and posterior probability for models from 
Table~\ref{tab:1}.}
\label{tab:5}
\begin{tabular}{c|ccccccccccc}
\hline
case &1&2&3&4&5&6&7&8a&8b&9&10 \\  
\hline
$\mathbf{prior}$&0.0909&0.0909&0.0909&0.0909&0.0909&0.0909&0.0909&0.0909&0.0909&0.0909&0.0909\\
$\mathbf{posterior}$&0.1315&0.0293&0.0240&0.3399&0.0561&0.0308&0.0025&0.0046&0.0048&0.0005&0.3757
\\
\hline
\end{tabular}
\end{table}

\begin{table}
\caption{Values of the prior and posterior probability for models from 
Table~\ref{tab:2}.}
\label{tab:6}
\begin{tabular}{c|cccccccccccc}
\hline
case &1&2&3&4&5&6a&6b&7&8&9&10a&10b \\  
\hline
$\mathbf{prior}$&0.083&0.083&0.083&0.083&0.083&0.083&0.083&0.083&0.083&0.083&0.083&0.083\\
$\mathbf{posterior}$&0.122&0.173&0.041&0.000&0.004&0.011&0.002&0.052&0.285&0.285&0.012&0.012\\
\hline
\end{tabular}
\end{table}

\begin{table}
\caption{Values of the prior and posterior probability for set of brane models 
from Table~\ref{tab:2}.}
\label{tab:7}
\begin{tabular}{c|cccccc}
\hline
case &2&3&4&5&6a&6b\\  
\hline
$\mathbf{prior}$&0.167&0.167&0.167&0.167&0.167&0.167\\
$\mathbf{posterior}$&0.749&0.175&0.019&0.000&0.047&0.008 \\
\hline
\end{tabular}
\end{table}

\begin{table}
\caption{Values of the prior and posterior probability for set of models from 
Table~\ref{tab:1}: $\{1, 4, 10\}$ and Table~\ref{tab:2}: $\{1, 2\}$.}
\label{tab:8}
\begin{tabular}{c|ccc|cc}
\hline
case &1&4&10&1&2\\  
\hline
$\mathbf{prior}$&0.20&0.20&0.20&0.20&0.20 \\
$\mathbf{posterior}$&0.13&0.34&0.38&0.06&0.08 \\
\hline
\end{tabular}
\end{table}

\section{Conclusion}
Main aim of the paper was exploring the Bayesian framework of model selection 
into discussion which cosmological model describe present accelerating phase 
expansion of the Universe. In principle there are two types of explanation why 
current Universe is accelerating. In the first group it is postulated existence 
of perfect fluid which violate the strong energy condition. The nature 
(Lagrangian) of such matter called dark energy is unknown although the 
cosmological constant is the most popular candidate for dark energy description.

The second group of explanations is based on hypothesis that dark energy could 
actually be the manifestation of a modification to the Friedmann equation 
arising from `new physics' (e.g. extra dimensions, generalized general 
relativity, etc.). Calculating the corrections to the standard FRW equation we 
can explore the phenomenology and different evolutional scenarios. Therefore 
there is no single hypothesis, instead there are several well-supported 
hypotheses (i.e. dark energy models) that are being entertained. It is just 
realization of concept of `Multiple Working Hypotheses' advocated by 
Chamberlin \cite{Chamberlin:1965}. In Tables~\ref{tab:1} and \ref{tab:2} we 
completed the 20 models in two classes which cover both types of explanation 
of acceleration of the Universe. Then we adopt the model selection 
methods to obtain `strength of evidence' comparison and ranking of candidate 
hypotheses of dark energy. Providing quantitative information to judge the 
`strength of evidence' is a key point of ranking models. The hypothesis 
testing (which only provides qualitative information significant vs. 
nonsignificant) is particularly limited in the model selection (for discussion 
of the likelihood based strength of evidence see Ref.~\cite{Royall:1997}).

Hence we obtain the set of models which are recommended by the AIC and Bayes factor. The results are following:
\begin{enumerate}
\item While the AIC recommended dark energy models which belong to the set 
$\{10, 4, 5, 6, 1, 2\}$ the Bayes factor is more restrictive because recommended models 
which are elements of subset $\{10, 4, 1\}$. This is due to different values of the coefficient in the definition of AIC and BIC quantities, so called penalty term for more complex models. In the AIC definition this coefficient is always equal $2$ whereas in the BIC definition it depends on simple size (here it is nearly equal $5$). 
\item Analogous recommendation can be performed for the class of models with 
a modified Friedmann equation, namely the AIC recommends a set of models 
$\{1, 8, 10a, 10b, 6a, 7\}$ and the Bayes factor favors class of models $\{2, 1\}$ 
which has non-empty intersection with models recommended by the AIC.  
\item One can construct also ranking within the best models in each category 
I and II by AIC and Bayes factor respectively. Then we obtain a set of models 
recommended by the AIC $\{10, 4, 5, 6, 1, 2\}_{I} \cup \{1\}_{II}$ and by 
the Bayes factor $\{10, 4, 1\}_{I}$.

Note that both quantities pointed out that model with noninteracting Chaplygin gas and baryonic matter 
is the best one from both sets of models under consideration. This is due 
to more a priori information about this model which we include in calculation. 
Two model parameters were fixed here: $\Omega_{\text{m},0}$ and $m$. In 
situation when both parameters are fitted the conclusion is changed. All 
additional information which we include in calculation are necessary and can 
change our final inference.

It should be point out that dark energy models better explain SNIa data than 
models with a modified Friedmann equation. 
\item We can perform ranking of the models within `brane paradigm' and then 
we obtain models $\{6a, 3, 2, 5\}$ recommended by the AIC and model  $\{2\}$ 
preferred by the Bayes factor.
\item One can conclude that the flat $\Lambda$CDM model (model indexed by $1$) 
has still substantial support with respect to better models. Note that the AIC 
indicates that there is no difference between  the $\Lambda$CDM $(k=0)$ model 
and the $\Lambda$CDM $(k \neq 0)$ model (indexed by $2$), both of them fit the 
data equally well. Whereas the Bayes factor denotes that there is an evidence 
against model $2$ with respect to model $1$.

Note that Bayes factor indicates that the flat $\Lambda$CDM model better 
explains SNIa data than models gathered in Table II.
\item For completeness we calculate posterior probability which measures 
probability for the model being the best one among the class of models under 
consideration (being the most favored model by data in hand). Then one can 
observe how prior believe about model probability change after inclusion data 
to analysis (see Tables \ref{tab:5}, \ref{tab:6}, \ref{tab:7}, \ref{tab:8}). 
Note that for all models recommended by 
the BIC posterior probability is greater than the prior one, whereas for models 
which do not belong to a preferred set of models the posterior is smaller than the 
prior.
\end{enumerate}   

\acknowledgments
Authors are grateful to dr W. Godlowski for fruitful discussion and the 
participants of the seminar on observational cosmology for comments. We also 
thank an anonymous referee for suggested improvements to this paper and 
additional references.

\end{document}